\documentstyle[12pt]{article}
\newcommand{\M}{{\cal M}}
\newcommand{\X}{{\cal X}}
\newcommand{\N}{{\cal N}}
\newcommand{\B}{{\cal B}}
\newcommand{\al}{\alpha}

\newcommand{\be}{\begin{equation}}
\newcommand{\ee}{\end{equation}}
\newcommand{\ba}{\begin{eqnarray}}
\newcommand{\ea}{\end{eqnarray}}
\newcommand{\ban}{\begin{eqnarray*}}
\newcommand{\ean}{\end{eqnarray*}}

\newcommand{\partderiv}[2]{\frac{\partial #1}{\partial #2}}

\newcommand{\e}{{\epsilon}}

\title{Classical and Quantum Equivalence Principle \\ in Terms of the 
        Path Group\footnote{Published in Helvetica Physica 
        Acta, 69, 301-304 (1996)}}
\author{Michael B. Mensky\\
P.N.Lebedev Physical Institute, 117924 Moscow, Russia}
\date{}

\begin{document}

\maketitle

\abstract{A natural mapping of paths in a curved space onto the paths 
in the corresponding (tangent) flat space may be used to reduce the 
curved-space-time path integral to the flat-space-time path integral. 
The dynamics of the particle in a curved space-time is expressed then 
in terms of an integral over paths in the flat (Minkowski) space-time. 
This may be called quantum equivalence principle. Contrary to the 
known DeWitt's definition of a curved-space path integral, the present 
definition leads to the covariant equation of motion without a scalar 
curvature term. The reduction of a curved-space path integral to the 
flat-space path integral may be expressed in terms of a representation 
of the path group. With the help of this representation all the 
results may be generalized to the case of an arbitrary external 
field.}

\section{Introduction}\label{intro}

The motion of a free classical point particle is described by a
direct line in the Minkowski space-time. The evolution of a
quantum point particle may be described by a path integral in
the Minkowski space. According to the Einstein's equivalence
principle, a point classical particle in a curved space-time
moves along a geodesic line. The motion of a quantum particle may
be described by an integral over paths in the curved space-time,
however the definition of such an integral is ambiguous. The
first definition given by B.DeWitt \cite{DeWitt} and some other
definitions lead to the equations of motions differing by the
coefficient in the term proportional to the scalar curvature $R$.

It will be pointed out in the present paper that the evolution of both 
classical and quantum particles in a curved space-time may be 
naturally described in terms of the Minkowski space-time. This 
description is based on the natural but non-holonomic mapping of 
curves in the curved space-time onto the curves in the Minkowski 
space-time. This results in the description of a classical particle 
moving in the curved space-time, by a direct line in the Minkowski 
space. The description of a quantum particle is given by the integral 
over paths in the Minkowski space, but with the corresponding operator 
in the integrand. This path-depending operator may be shown to be a 
representation of the path group.

\section{Mapping of curves and the equivalence
principle}\label{mapping}

It is well known that there is no natural point-to-point mapping
of a curved space onto the Minkowski space. However a natural
correspondence exists between the curves in the Minkowski space and
the curves in the curved space provided the starting point of
these curves and (local) reference frame in this point are
fixed.

To describe this correspondence, let us identify the Minkowski 
space $\M$ with a tangent space $\M_x$ to the curved 
(pseudo-Riemannian) space $\X$ in the starting point $x$ of the 
curves, and the reference frame of the Minkowski space with some 
orthonormal local frame $n=\{ n_{\al}\in\M_x: \al=0,1,2,3\}$ in 
the specified point $x$.

1. Let $[x]=\{ x(\tau')\in\X: 0\le\tau'\le\tau\}$ be a curve in
$\X$ starting in $x$ and $n(\tau')$ a result of parallel 
transport of $n$ along the curve $[x]$. Then the curve $[\xi]=\{ 
\xi(\tau')\in\M: 0\le\tau'\le\tau\}$ in $\M$ (with $\xi(0)$ 
coinciding with the origin of $\M$) may be defined unambiguously 
by the condition that its tangent vector $\dot{\xi}(\tau')$, 
being expanded in respect to the reference frame of the Minkowski 
space, has the same coefficients as the tangent vector 
$\dot{x}(\tau')$ expanded in respect to the local frame 
$n(\tau')$.

As a result, the natural correspondence of curves in $\M$
(starting in the origin) with the curves in $\X$ (starting in $x$)
is established, provided the local frame $n$ in the point $x$ is
given. Instead of the curve in $\M$ with the fixed starting
point, $[\xi]$ may be understood as a class of curves differing
by general translation, $[\xi]\sim[\xi']$, if
$\xi'(\tau')=\xi(\tau')+a$.

Naturalness of the mapping defined in this way may be formulated
as follows: If the observer located in the space-time point $x$
has to describe the path $[x]$ (in the curved space-time $\X$) as
a line in his ``flat map'' (with the geometry of the Minkowski
space $\M$), he should choose the curve $[\xi]$. Such a ``flat
modelling'' is defined for an arbitrary curve $[x]$ in $\X$. 
Geodesic lines $[x]$ will be associated then with direct lines
$[\xi]$. This gives a simple formulation of the Einstein's
equivalence principle \cite{flat-sym}: local ``flat models'' for 
trajectories of a point particle are direct lines. 

2. By the described procedure, a class of curves $[\xi]$
in $\M$ and an orthonormal local frame $n\in\N$ ($\N$ being a
fiber bundle of all orthonormal local frames) determine one more
local frame $n[\xi]\in\N$ by the rule: if $n=n(0)$, then
$n[\xi]=n(\tau)$ in the notations introduced earlier. Therefore,
each class of curves $[\xi]$ determines a mapping of $\N$ onto
itself.

This mapping may be described in a more formal way \cite{TMF74} with
the help of the concept of the basis vector fields in the
fiber bundle of local frames. To introduce this concept, it is
convenient to consider the fiber bundle $\B$ of all local frames
over $\X$. The coordinates $x^{\mu}$ of the points in $\X$ and
the components $b_{\al}^{\mu}$ of the vectors of the local frame
$b$ in this point may serve as coordinates of the manifold $\B$.
Then the basis vector fields $B_{\al}$ are following:
\be\label{Ba}
B_{\al}
=b_{\al}^{\mu}\partderiv{}{x^{\mu}}
+b_{\al}^{\mu}b_{\beta}^{\nu}\Gamma_{\mu\nu}^{\lambda}(x)
  \partderiv{}{b_{\beta}^{\lambda}}.
\ee

The vector fields $B_{\al}$ are horizontal in the fiber bundle
$\B$ and their restrictions on the fiber bundle $\N$ of
orthonormal local frames are horizontal in $\N$. We shall need
these restrictions rather than the complete fields $B_{\al}$. For
simplicity, we shall denote them by the same letters.

The horizontal vector fields allow one to define, as an ordered
exponential of an integral, the following
operator acting in the space of functions on $\N$:
\be\label{V}
V[\xi]=Pe^{\int B_{\al}d\xi^{\al}}
  = \lim_{N\to\infty}e^{B_{\al}\Delta\xi_N^{\al}}
  \dots e^{B_{\al}\Delta\xi_1^{\al}}. 
\ee
The set of operators $V[\xi]$ forms a representation of the path 
group \cite{PathGr} (generalizing translations). 
In terms of these operators, the mapping
$n\rightarrow n[\xi]$ may be defined as follows:
\be\label{Vpsi}
(V[\xi]\Psi)(n)=\Psi(n[\xi]).
\ee

\section{Quantum equivalence principle}\label{sec-equiv}

Evolution of a quantum particle is described by the propagator
which may be expressed in the form of a path integral. The path
integral in a curved space-time may be reduced to the path
integral in the Minkowski space-time but with the operator
$V[\xi]$ in the integrand.

1. To do this, let us describe states of the particle by
functions $\Psi(n)$ on the fiber bundle of (orthonormal) local
frames instead of usual functions $\psi(x)$ on the space-time
\cite{TMF74}.
Both functions are connected in a very simple way for
the scalar particle:
$$
\Psi(n)=\psi(x), \quad x=\pi(n)
$$
where the canonical projection $\pi:\, \N\to \X$ associates the
point $x$ with the local frame $n$ in this point. This definition 
may be naturally generalized onto the case of a spinning 
particle: 
$$
\Psi(n\lambda)=D(\lambda^{-1})\,\Psi(n), \quad
\psi(x)=\Psi(\sigma(n)), \quad x=\pi(n).
$$
Here $\sigma$ is an arbitrary section of the fiber bundle $\N$, 
and $D$ a representation of the Lorentz group describing spin
of the particle.

2.The evolution of a quantum particle may be described by a
propagator $U(x,x')$, but we shall use the corresponding operator 
$U$ (for which the propagator is a kernel). In the Minkowski 
space-time this operator may be expressed in the form of a path 
integral 
\be\label{evol-operator}
U=\int_0^{\infty} d\tau\, \e^{-im^2 \tau}U_{\tau}, \quad
U_{\tau}=\int d[\xi]\, \e^{(-i/4)\int_0^{\tau}d\tau\,
\dot{\xi}^{\alpha}\dot{\xi}_{\alpha}} \, V[\xi]
\ee
where $V[\xi]$ is an operator of displacement along the path
$[\xi]$:
$$
(V[\xi] \psi)(\xi)=\psi(\xi-\Delta\xi), \quad
\Delta\xi = \xi(\tau)-\xi(0).
$$

The propagator in a curved space-time may be defined \cite{TMF74} by 
the same formulas (\ref{evol-operator}) but with the expression 
(\ref{V}) for the operator $V[\xi]$. The dynamics of the particle in a 
curved space-time is expressed then in terms of the integral over 
paths in the flat space-time (Minkowski space). This may be called 
{\em quantum equivalence principle}. The resulting definition of the 
curved-space path integral differs from the known DeWitt's definition 
\cite{DeWitt} in that it leads to the covariant equation of motion 
with no term proportional to the scalar curvature. An essentially 
equivalent though apparently different definition of the path integral 
in a curved space has been given in \cite{Kleinert}.

The operator $V[\xi]$ which maps flat-space paths onto the
curved-space paths may be shown to form a representation of the 
path group \cite{PathGr}. With the help of this representation all 
the results may be generalized on the case of an arbitrary 
external field (gauge, gravitational or gauge plus gravitational 
fields).

\centerline{* * *}

This work was supported in part by the Russian Foundation for 
Basic Research, grant 95-01-00111a.

\end{document}